\documentclass[aps,prd,showpacs,superscriptaddress]{revtex4}
\usepackage{epsfig}
\newcommand{\ben}{\begin{eqnarray}}
\newcommand{\een}{\end{eqnarray}}
\newcommand{\nnu}{\nonumber\\}
\newcommand{\bef}{\begin{figure}[htb]\centering}
\newcommand{\eef}{\end{figure}}
\newcommand{\tq}{T_{q, F}}
\newcommand{\td}{T_{\Delta{q}, F}}
\newcommand{\sla}[1]{\slash\!\!\!{#1}}

\begin{document}
\title{Quark-gluon correlation functions relevant to single transverse spin asymmetries}

\date{\today}

\author{Zhong-Bo Kang}
\email{zkang@bnl.gov}
\affiliation{RIKEN BNL Research Center,
                Brookhaven National Laboratory,
                Upton, NY 11973-5000}
                
\author{Jian-Wei Qiu}
\email{jqiu@bnl.gov}
\affiliation{Physics Department,
                Brookhaven National Laboratory,
                Upton, NY 11973-5000}
\affiliation{C.N. Yang Institute for Theoretical Physics,
                Stony Brook University,
                Stony Brook, NY 11794-3840} 
\affiliation{Department of Physics and Astronomy,
                Iowa State University,
                Ames, IA 50011}   
                                
\author{Hong Zhang}
\email{hongzh87@iastate.edu}
\affiliation{Department of Physics and Astronomy,
                Iowa State University,
                Ames, IA 50011}

\begin{abstract}
We investigate the relative size of various twist-3 quark-gluon correlation functions relevant to single transverse spin asymmetries (SSAs) in a quark-diquark model of the nucleon. We calculate the quark-gluon correlation function $T_{q,F}(x, x)$ that is responsible for the gluonic pole contribution to the SSAs, as well as $T_{q,F}(0, x)$ and $T_{\Delta{q}, F}(0, x)$ responsible for the fermionic pole contributions. We find in both cases of a scalar diquark and an axial-vector diquark that at the first nontrivial order only the $\tq(x, x)$ is finite while all other quark-gluon correlation functions vanish. Using the same model, we evaluate quark Sivers function and discuss its relation to the $\tq(x, x)$. We also discuss the implication of our finding to the phenomenological studies of the SSAs.
\end{abstract}

\maketitle

%%%%%%%%%%%%%%%%%%%%%%
\section{Introduction}

The novel phenomenon of single transverse-spin asymmetry (SSA),
$A_N \equiv (\sigma(\vec{S}_\perp)-\sigma(-\vec{S}_\perp))
/(\sigma(\vec{S}_\perp)+\sigma(-\vec{S}_\perp))$, 
defined as the ratio of the difference and the sum of the cross sections when the single transverse spin vector $\vec{S}_\perp$ is flipped, was first observed in the hadronic $\Lambda^0$ production at Fermilab in 1976 as a surprise \cite{Bunce:1976yb}. Large SSAs, as large as 30 percent, have been consistently observed in various experiments involving one polarized hadron at different collision energies, and have attracted tremendous interest from both experimental and theoretical sides in recent years \cite{D'Alesio:2007jt}. The size of the observed SSAs presented a challenge to the early QCD calculation \cite{Kane:1978nd}. As a consequence of the parity and time-reversal invariance of the strong interaction, the SSAs in high energy collisions are directly connected to the transverse motion of quarks and gluons inside the transversely polarized hadron. The measurement of SSAs provides an excellent opportunity to probe a new domain of QCD dynamics. The understanding of the physics behind the measured asymmetries should have a profound impact on our knowledge of strong interaction and hadron structure.

Two complementary QCD-based approaches have been proposed to analyze and to explore the physics behind the measured SSAs: the transverse momentum dependent (TMD) factorization approach \cite{Siv90,Collins93,Brodsky,MulTanBoe,TMD-dis,boermulders,mulders} and the collinear factorization approach \cite{Efremov,qiu_spin,qiufermion,koike,Qiu:2007ar}. Both approaches have been applied extensively to phenomenological studies \cite{Bacchetta:2007sz,Collins:2002kn, Kang:2009bp, Kang:2009sm,Boer:2009nc,Kang:2010pr,Kouvaris:2006zy,Kang:2008qh,Koike:2009ge}. The TMD factorization approach is more suitable for evaluating the SSAs of scattering processes with two very different momentum transfers, $Q_1\gg Q_2 \gtrsim \Lambda_{\rm QCD}$. Having one large observed scale, $Q_1\gg \Lambda_{\rm QCD}$, is necessary for using perturbative QCD and the TMD factorization approach although it is not sufficient \cite{Collins:2007nk,Vogelsang:2007jk,Collins:2007jp,Rogers:2010dm}. For observables for which the TMD factorization is valid, this approach has an advantage for directly probing active parton's transverse motion at the scale, ${\cal O}(Q_2)$, inside a polarized hadron in the form of TMD parton distribution functions (PDFs). On the other hand, the collinear factorization approach is more relevant to the SSAs of scattering cross sections with all observed momentum transfers $Q\gg \Lambda_{\rm QCD}$. In the QCD collinear factorization approach, the leading power contribution to the cross sections in the $1/Q$ expansion cancels in evaluating the asymmetry because of the parity and time-reversal invariance of the theory. Therefore, the asymmetry directly probes the correlation of quarks and gluons inside a polarized nucleon in the form of the twist-3 quark-gluon and tri-gluon correlation functions \cite{qiu_spin}. Although the two approaches each have their own kinematic domain of validity, they describe the same physics and are consistent with each other in the regime where they both apply \cite{UnifySSA, Bacchetta:2008xw}.  

In both TMD and collinear factorization approaches of QCD, the size of calculated SSAs is proportional to some nonperturbative functions:\  the TMD PDFs and the twist-3 three-parton correlation functions, respectively. The predictive power of both approaches relies on the validity of the respective factorization and the knowledge of these nonperturbative functions \cite{JiMaYu04,Qiu_sterman}. QCD perturbation theory could be used to study the quantum evolution of these functions from one perturbative scale to another where these functions were probed \cite{kqevo,
Zhou:2008mz,Vogelsang:2009pj,Braun:2009mi}. But, the absolute normalizations of these functions or the boundary conditions - the input functions for solving the evolution equations have to be extracted from data of measured asymmetries.  With the recent measurements of SSAs \cite{HERMES,COMPASS,SSA-rhic}, we have gained valuable information on the TMD PDFs \cite{Anselmino:2008sga} and twist-3 correlation functions \cite{Kouvaris:2006zy}. Although precise data from future experiments could certainly help fix these nonperturbative functions, a good model calculation of these unknown functions could provide important insight into the mechanisms for generating the observed novel asymmetries and valuable guideline to the relative importance and size of various functions. In this paper, we present our calculations of all twist-3 quark-gluon correlation functions relevant to the SSAs in the collinear factorization approach in a quark-diquark model of the nucleon \cite{Brodsky,Bacchetta:2008af}. We calculate these quark-gluon correlation functions with both scalar and axial-vector diquarks. We also evaluate in the same model the quark Sivers function, the spin dependent part of the TMD quark distribution, and discuss its relation to the twist-3 quark-gluon correlation function, $\tq(x, x)$ \cite{boermulders}.

In order to generate a nonvanishing SSA in high energy hadronic collisions, one needs to generate a parton-level spin flip and a phase difference between the scattering amplitude and its complex conjugate. In the QCD collinear factorization approach to the SSAs, the spin flip at the hard collision was achieved by the interference between an active single parton state and an active two-parton composite state of the scattering amplitude; and the phase difference was generated by the interference between the real part and the imaginary part of the short-distance partonic scattering amplitude \cite{qiu_spin}. We obtain the leading contribution to the imaginary part of the partonic scattering amplitude by taking the unpinched pole of the partonic scattering amplitude \cite{Efremov,qiu_spin}.  It is the interference between the single active parton state and the two-parton composite state that requires the calculated SSAs to be proportional to the twist-3 quark-gluon correction functions, $\tq(x_1, x_2)$ and $\td(x_1, x_2)$, and tri-gluon correlation functions, $T_{G,F}^{(f,d)}(x_1,x_2)$ and $T_{\Delta G,F}^{(f,d)}(x_1,x_2)$, convoluted with corresponding partonic scattering through two independent momentum fractions of the three active partons,  $x_1$ and $x_2$ \cite{kqevo,Braun:2009mi}.  Taking the pole of the parton scattering amplitude effectively fixes one of the two momentum fractions.  Depending on the number of observed hard momentum scales, the partonic scattering amplitude has different pole structure.  For cross sections with a single observed hard scale, such as $p_T$ of single inclusive pion production in hadronic collisions, the leading pole contribution is from taking the residue of the pole, which is effectively setting the momentum fraction of one of the three active partons to zero \cite{Efremov,qiu_spin}.  This contribution is often referred as the soft-pole contribution. The so-called gluonic (or fermionic) pole contribution refers to the situation when the active gluon (or (anti)quark) momentum fraction was set to zero.  For cross sections with more than one observed hard scale, such as inclusive pion production in lepton-hadron deep inelastic scattering when both pion momentum $p_T$ and virtual photon invariant mass $Q$ are large, the leading pole contribution could also come from the situation when all active parton momentum fractions are finite, known as the hard-pole contribution \cite{Luo:1994np, Guo:1997it}.  In this paper, we present the model calculation only for quark-gluon correlation functions corresponding to the soft-pole contribution to the SSAs,  which include $\tq(x,x)$ and $\td(x,x)$ for the gluonic pole contribution, and $\tq(0,x)$, $\td(0,x)$, $\tq(x,0)$, and $\td(x,0)$ for the fermionic pole contribution.  

In general, the calculated SSAs of cross sections with one observed hard momentum scale receive contributions from both the gluonic and fermionic poles, and the partonic hard parts for these two contributions often have the similar size \cite{Kouvaris:2006zy,Koike:2009ge}.  
On the other hand, the quark-gluon correlation functions corresponding to the gluonic pole and fermionic pole contribution represent very different dynamical structure inside the polarized proton.  For the gluonic pole contribution, the quark-gluon correlation functions, $\tq(x,x)$ and $\td(x,x)$, represent the quantum interference between a quark state of momentum fraction $x$ and a quark-gluon composite state of the same momentum fraction with the quark carrying all the momentum fraction $x$, while for the fermionic pole contribution, the total momentum fraction of the quark-gluon composite state is carried by the gluon.  The relative size of these two types of quark-gluon correlation functions certainly provides interesting information on the dynamical structure of a polarized proton.

In terms of the simple quark-diquark model of the nucleon \cite{Brodsky,Bacchetta:2008af}, we find in both cases of a scalar diquark and an axial-vector diquark that at the first non-trivial order, {\it all} quark-gluon correlation functions corresponding to the fermionic pole contribution,  $\tq(0,x)$, $\td(0,x)$, $\tq(x,0)$, and $\td(x,0)$, vanish.  For those functions corresponding to the gluonic pole contribution, $\tq(x,x)$ is finite while $\td(x,x)=0$ which is consistent with the result of symmetry argument \cite{kqevo}.  Our results, although from a model calculation, indicate that the fermionic pole contribution to the SSAs is likely to be less important than the gluonic pole contribution.  This conclusion seems to be consistent with a general expectation that a quark-gluon state with the quark carrying all of its momentum is more likely than a quark-gluon state with the gluon carrying all of its momentum to interfere with a quark state of the same momentum \cite{qiu_spin}. Our finding could help streamline the phenomenological studies of the SSAs by starting with a much smaller number of nonperturbative twist-3 correlation functions.

The rest of our paper is organized as follows. In Sec.~\ref{allfun}, we introduce the operator definition of all twist-3 quark-gluon correlation functions and discuss their symmetry properties.  In Sec.~\ref{model}, we introduce the quark-diquark model of the nucleon and its Feynman rules for both cases of a scalar diquark and an axial-vector diquark, and present our calculations for the twist-3 quark-gluon correlation functions relevant to both gluonic and fermionic pole contributions to the SSAs. In terms of the same nucleon model, we calculate the quark Sivers functions in Sec.~\ref{conn}, and discuss the connection between the twist-3 quark-gluon correlation functions and the TMD parton distribution functions. Finally, we give our summary and conclusions in Sec.~\ref{sum}.

%%%%%%%%%%%%%%%%%%%%%%
\section{The twist-3 quark-gluon correlation functions}
\label{allfun}

The twist-3 three-parton correlation functions in the QCD collinear factorization approach to the SSAs could be represented by the cut forward scattering diagram in Fig.~\ref{feytwist3} with proper cut vertices \cite{kqevo}.  These correlation functions measure the net effect of quantum interference between two scattering amplitudes of the transversely polarized proton:\ one with a single active parton and the other with two active partons, participating in the short-distance hard scattering.
\bef
\psfig{file=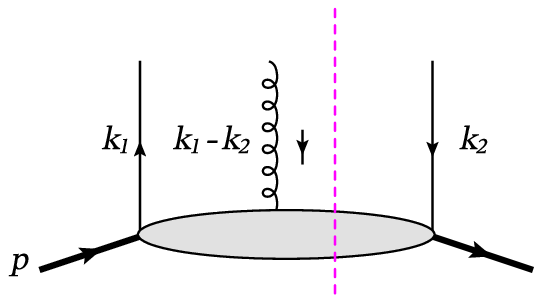, width=2.1in}
\caption{The Feynman diagram representation for the twist-3 quark-gluon correlation functions, where $k_i\approx x_ip$ with $i=1,2$.  Contracting the three active partons with different cut vertices leads to different quark-gluon correlation functions \cite{kqevo}.}
\label{feytwist3}
\eef
A complete set of twist-3 three-parton correlation functions relevant to the SSAs has been constructed in Refs.~\cite{kqevo,Braun:2009mi}, which includes two independent quark-gluon correlation functions, $\tq(x_1, x_2)$ and $\td(x_1, x_2)$.  They could be derived from the following quark-gluon matrix element of a transversely polarized hadron \cite{koike},
\ben
{\mathcal M}^\sigma(x_1, x_2,s_T)
&=&
\int\frac{dy_1^- dy_2^-}{2\pi}e^{ix_1p^+y_1^-+i(x_2-x_1)p^+y_2}
\langle p,s_T|\bar{\psi}_q(0)g F^{\sigma +}(y_2^-)\psi_q(y_1^-)|p,s_T\rangle
\nnu
&=&
\frac{1}{2}\left[\,\sla{\bar{n}}\,\epsilon^{\sigma s_Tn\bar n} \tq(x_1, x_2)
+\gamma^5 \,\sla{\bar{n}}  \,i s_T^\sigma \td(x_1, x_2)+\cdots \right],
\label{matrix}
\een
where the proper gauge links have been suppressed and $x_i = k_i\cdot n/p\cdot n$ with $i=1,2$ are two independent parton momentum fractions.  $\bar{n}^\mu=[1^+, 0^-, 0_\perp]$ and $n^\mu=[0^+, 1^-, 0_\perp]$ are two light-like vectors with $\bar{n}\cdot n=1$, and the ellipsis represents terms at twist-four and higher. From Eq.~(\ref{matrix}), we derive the expressions for the relevant quark-gluon correlation functions \cite{kqevo},
\ben
\tq(x_1, x_2)&=&\int\frac{dy_1^- dy_2^-}{4\pi}e^{ix_1p^+y_1^-+i(x_2-x_1)p^+y_2}
\langle p,s_T|\bar{\psi}_q(0)\gamma^+\left[ \epsilon^{s_T\sigma n\bar{n}} g F_\sigma^{~ +}(y_2^-)\right] \psi_q(y_1^-)|p,s_T\rangle\, ,
\label{Tq}
\\
\td(x_1, x_2)&=&\int\frac{dy_1^- dy_2^-}{4\pi}e^{ix_1p^+y_1^-+i(x_2-x_1)p^+y_2}
\langle p,s_T|\bar{\psi}_q(0)\gamma^+\gamma^5\left[i s_T^\sigma g F_\sigma^{~ +}(y_2^-)\right] \psi_q(y_1^-)|p,s_T\rangle \, .
\label{Tdq}
\een
From parity and time-reversal invariance, these two functions have the following symmetry property under the exchange of the two arguments $x_1\leftrightarrow x_2$ \cite{qiu_spin, kqevo}:
\ben
\tq(x_1, x_2)=\tq(x_2, x_1),
\qquad
\td(x_1, x_2)=-\td(x_2, x_1).
\label{tqdq}
\een
The leading order gluonic pole contribution to the SSAs is connected to the diagonal quark-gluon correlation functions, $\tq(x, x)$ and $\td(x, x)$ \cite{Efremov,qiu_spin}.  On the other hand, the leading order fermionic pole contribution to the SSAs is connected to the off-diagonal quark-gluon correlation functions, $\tq(0,x)$ and $\td(0,x)$, or $\tq(x,0)$ and $\td(x,0)$ \cite{Efremov,qiu_spin}.  From Eq.~(\ref{tqdq}), we have $\td(x, x)=0$.  In the next section, we calculate these correlation functions in the quark-diquark model of the nucleon, and test the symmetry properties in Eq.~(\ref{tqdq}).

%%%%%%%%%%%%%%%%%%%%%%
\section{Model calculation of twist-3 quark-gluon correlation functions}
\label{model}

In this section, we calculate the twist-3 quark-gluon correlation functions relevant to the gluonic and fermionic pole contributions to the SSAs in the quark-diquark model of the nucleon \cite{Brodsky,Bacchetta:2008af}.  We consider two possible situations in which the spectator diquark is either a scalar particle or an axial-vector particle. 

%%%%%%%%%%%%%%%%%%%%%%
\subsection{The quark-diquark model of the nucleon}

In the quark-diquark model of the nucleon \cite{Brodsky,Bacchetta:2008af}, the nucleon of mass $M$ consists of a constituent quark 
of mass $m$ and a diquark spectator of mass $M_s$.  
\bef
\psfig{file=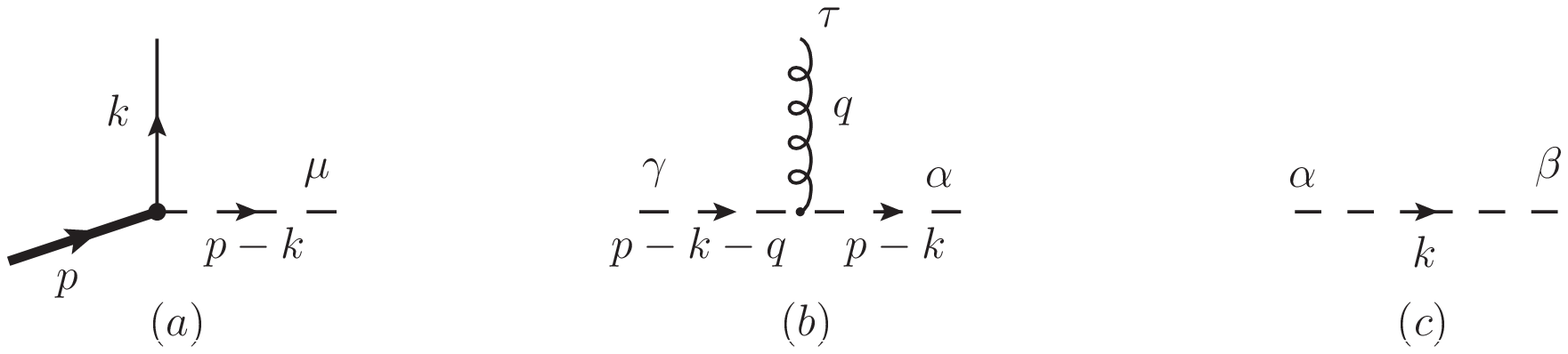, width=4.5in}
\caption{Feynman diagrams to define the Feynman rules in the quark-diquark model of the nucleon: (a) vertex links the nucleon, the quark, and the diquark, (b) interaction vertex between the gluon and the diquark, and (c) the diquark propagator. The diquark could be a scalar particle or an axial-vector particle.  The Lorentz indices are for the gluon and the axial-vector diquark.}
\label{feynrule}
\eef
The interaction between the nucleon, the quark, and the diquark is given by the following Feynman rule for the vertex in Fig.~\ref{feynrule}(a),
\ben
&i\lambda_s\, F_s(k^2)    
&\qquad \mbox{scalar diquark,}
\\
&i\frac{\lambda_v}{\sqrt{2}}\gamma^\mu \gamma^5\, F_v(k^2)
& \qquad  \mbox{axial-vector diquark,}
\een
where $\lambda_{s,v}$ represent the point-like interaction strength with subscripts $s$ and $v$ for a scalar and an axial-vector diquark, respectively, $F_{s,v}(k^2)$ are suitable form factors as a function of $k^2$ - invariant mass square of the constituent quark.  $F_{s,v}(k^2)=1$ is for a point-like vertex interaction.  As explained later, a properly chosen form factor could help control the ultraviolet behavior of the calculated quark-gluon correlation functions.  The Feynman rule for the coupling between the gluon and the diquark in Fig.~\ref{feynrule}(b) is given by 
\ben
&i\, g_s\, (2p-2k-q)^\tau 
&\qquad \mbox{scalar diquark,}
\\
&i\, g_v\, V^{\tau\gamma\alpha}(q, p-k-q, k-p)
&\qquad \mbox{axial-vector diquark,}
\een
with the coupling strength $g_s$ and $g_v$ for a scalar and an axial-vector diquark, respectively.  Here $V^{\tau\gamma\alpha}(q, p-k-q, k-p)$ is given by \cite{Bacchetta:2008af}
\ben
V^{\tau\gamma\alpha}(q, p-k-q, k-p)=
g^{\tau\gamma}(2q-p+k)^\alpha
+g^{\gamma\alpha}(2p-2k-q)^\tau
+g^{\alpha\tau}(k-p-q)^\gamma\, .
\een
The Feynman rule for a scalar diquark propagator is the same as that of a normal scalar particle, while the Feynman rule for the axial-vector diquark propagator in Fig.~\ref{feynrule}(c) is given by 
\ben
\frac{i}{k^2-M_s^2}\, d^{\alpha\beta}(k,M_s),
\een
where the polarization tensor $d^{\alpha\beta}(k,M_s)$ has the following form \cite{Bacchetta:2008af}
\ben
d^{\alpha\beta}(k,M_s)=-g^{\alpha\beta}+\frac{k^\alpha n^\beta+k^\beta n^\alpha}{n\cdot k}-\frac{M_s^2n^\alpha n^\beta}{(n\cdot k)^2}\, ,
\een
which has the property, $k_\alpha\, d^{\alpha\beta}(k,M_s) = 0$ when $k^2=M_s^2$.

As we will show below, the twist-3 quark-gluon correlation functions calculated with the point-like coupling between the nucleon, the quark, and the spectator diquark, $F_{s,v}(k^2)=1$, have logarithmic ultraviolet divergences when $k^2\to \infty$.  Since we are mainly interested in the long-distance behavior of the quark-gluon correlation functions, we could choose a proper form factor to eliminate the divergence from the region of phase space where $k^2 \gg M^2$, the mass scale of the nucleon, while preserve the dynamics at $k^2 \sim M^2$.  Several choices for the form factor were introduced and discussed in Ref.~\cite{Bacchetta:2008af}.  In our calculation below, we assume that the form factor for a scalar diquark is the same as that for an axial-vector diquark, and choose a dipolar form factor \cite{Bacchetta:2008af}
\ben
F(k^2)=F_{s}(k^2)=F_{v}(k^2) = \frac{k^2-m^2}{\left[k^2-\Lambda_s^2\right]^2}\, \Lambda_{s}^2 \, ,
\label{dipolar}
\een
where $\Lambda_{s}^2 \gtrsim M^2$ is an ultraviolet cutoff.  Note that in Eq.~(\ref{dipolar}) we multiplied the dipolar form factor in Ref.~\cite{Bacchetta:2008af} by an extra $\Lambda_{s}^2$ so that the form factor has the right dimension.  Such a difference by a constant factor does not affect any of our conclusions derived below.   We will also demonstrate below that the introduction of this form factor smoothly suppresses the influence of the ultraviolet region of $k_\perp^2$ or $k^2$ without affecting the main conclusions of this paper.

%%%%%%%%%%%%%%%%%%%%%%
\subsection{Calculation with a scalar diquark}

All quark-gluon correlation functions could be represented by the same cut forward scattering diagram in Fig.~\ref{feytwist3}.  The difference of various quark-gluon correlation functions is from the difference in cut vertices contracted to the three active partons in the diagram \cite{kqevo}.  The form of cut vertices for both $\tq(x_1,x_2)$ and $\td(x_1,x_2)$ as well as the tri-gluon correlation functions is gauge dependent and was derived in Ref.~\cite{kqevo}.  In this paper, we work in the light-cone gauge.  For $x_1\equiv x+y$ and $x_2\equiv x$, the cut vertices are given by \cite{kqevo}
\ben
{\cal V}_{q,F}^{\rm LC}
&=&
\frac{\gamma^+}{2p^+}\,2\pi g\delta\left(x-\frac{k^+}{p^+}\right)
y\,\delta\left(y-\frac{q^+}{p^+}\right)
\left(i\,\epsilon^{s_T\mu n\bar{n}}\right)
\left[-g_{\mu\sigma}\right]\,
{\cal C}_q \, ,
\label{qcut}
\\
{\cal V}_{\Delta q,F}^{\rm LC}
&=&
\frac{\gamma^+\gamma^5}{2p^+}\,2\pi g\delta\left(x-\frac{k^+}{p^+}\right)
y\, \delta\left(y-\frac{q^+}{p^+}\right)
\left(-s_T^\mu \right)\,
\left[-g_{\mu\sigma}\right]\, {\cal C}_q
\label{dcut}
\een
where $g$ with $g^2=4\pi\alpha_s$ is the strong coupling constant included in the definition in Eq.~(\ref{matrix}), ${\cal C}_q$ is the fermionic color contraction factor given by \cite{kqevo}
\ben
\left({\cal C}_q\right)^c_{ij} 
= \left(t^c\right)_{ij} \, ,
\label{color_qc}
\een
with quark and gluon color indices, $i,j=1,2,3=N_c$ and $c=1,2,...,8=N_c^2-1$, respectively, and $t^c$ are the generators of the fundamental representation of $SU(3$) color. 
\bef
\psfig{file=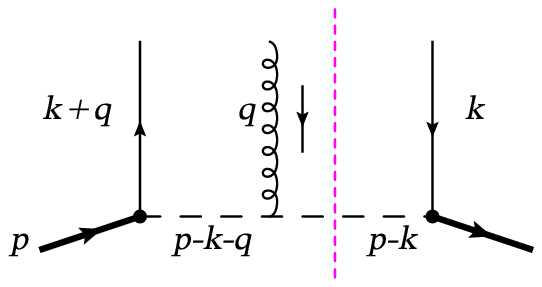, width=2.1in}
\caption{The lowest order Feynman diagram for twist-3 quark-gluon correlation functions in the quark-diquark model.}
\label{tffey}
\eef

The contribution to the twist-3 quark-gluon correlation functions $\tq(x+y, x)$ and $\td(x+y, x)$ at the lowest non-trivial order is given by the Feynman diagram in Fig.~\ref{tffey}. We first study these correlation functions with a scalar diquark. Applying the cut vertex in Eq.~(\ref{qcut}) to the diagram in Fig.~\ref{tffey}, we obtain
\ben
\tq^{(s)}(x+y, x)
&=&
-N_c C_F \frac{g\,\lambda_s^2\, g_s\, \pi^2}{p^+}
\int\frac{d^4k}{(2\pi)^4}\frac{d^4q}{(2\pi)^4}\,
\delta\left(x-\frac{k^+}{p^+}\right)y\, \delta\left(y-\frac{q^+}{p^+}\right)
\delta\left((p-k)^2-M_s^2\right)
\nnu
&&
\times
\epsilon^{s_T\sigma n\bar n}(2p-2k-q)^\tau d_{\sigma\tau}(q)\,
{\rm Tr}\left[\gamma^+(\sla{k}+\sla{q}+m)(\sla{p}+M)
\gamma^5 \sla{s_T}(\sla{k}+m)\right]
\nnu
&&
\times
\frac{1}{k^2-m^2-i\epsilon}
\frac{1}{q^2+i\epsilon}
\frac{1}{(k+q)^2-m^2+i\epsilon}
\frac{1}{(p-k-q)^2-M_s^2+i\epsilon}
F(k^2) F((k+q)^2)\, ,
\label{tqf}
\een
where $\epsilon=0^+$ represents a small positive parameter, the superscript $(s)$ indicates the scalar diquark, and the gluon polarization tensor $d_{\sigma\tau}(q)$ is given by
\ben
d_{\sigma\tau}(q)=-g_{\sigma\tau}+\frac{q_\sigma n_\tau+q_\tau n_\sigma}{q\cdot n}\, .
\een
Performing the integration over $k^+$, $k^-$, and $q^+$ by using the three $\delta$-functions in 
Eq.~(\ref{tqf}), we obtain
\ben
\tq^{(s)}(x+y, x)
&=&
-N_c C_F \frac{g\, \lambda_s^2\, g_s}{16 \pi p^+}
\int\frac{d^2q_\perp}{(2\pi)^2}
\frac{d^2k_\perp}{(2\pi)^2}
\frac{1}{k_\perp^2+L_s^2(m^2)}
\int \frac{dq^-}{2\pi}
\nnu
&&
\times
\epsilon^{s_T\sigma n\bar n}(2p-2k-q)^\tau (q^+ g_{\sigma\tau}-q_\sigma n_\tau)
{\rm Tr}\left[\gamma^+(\sla{k}+\sla{q}+m)(\sla{p}+M)
\gamma^5 \sla{s_T}(\sla{k}+m)\right]
\nnu
&&
\times
\frac{1}{q^2+i\epsilon}
\frac{1}{(k+q)^2-m^2+i\epsilon}
\frac{1}{(p-k-q)^2-M_s^2+i\epsilon}\,
F(k^2)\, F((k+q)^2)\, ,
\label{tqf1}
\een 
where 
\ben
k^2=m^2-\frac{1}{1-x}\left[k_\perp^2+L_s^2(m^2)\right]
\label{k2}
\een
with $L_s^2(m^2)$ given by
\ben
L_s^2(m^2)=xM_s^2+(1-x)m^2-x(1-x)M^2
\label{Ls2}
\een
independent of $q^-$.

The integration over $q^-$ is crucial and is done by taking the residue of relevant pole(s) of the integrand in Eq.~(\ref{tqf1}), which provides the necessary phase for a real quark-gluon correlation function $\tq(x+y,x)$.  Since we are interested in the leading gluonic and fermionic pole contribution to the SSAs, we exam below the pole structure of the integrand in Eq.~(\ref{tqf1}) at $y=0$ (gluonic pole) and $x+y=0$ (fermionic pole) while $x>0$.  From 
\ben
(p-k-q)^2-M_s^2+i\epsilon
=-2(1-x-y)p^+ q^{-} - \frac{y(k_\perp^2+M_s^2)}{1-x}-2k_\perp\cdot q_\perp -q_\perp^2+i\epsilon = 0\, ,
\een
and $x+y<1$, we derive the location of the corresponding pole at
\ben
q^-=-\frac{1}{2(1-x-y)p^+}\left[\frac{y(k_\perp^2+M_s^2)}{1-x}+2k_\perp\cdot q_\perp
+q_\perp^2\right]+i\epsilon\, ,
\een
which is in the upper half plane of the $q^-$.  This pole survives and stays in the upper half plane at both limits: $y=0$ (gluonic pole) and $x+y=0$ (fermionic pole). However, the potential poles from $q^2+i\epsilon=0$ and $(k+q)^2-m^2+i\epsilon=0$ are sensitive to these two limits.  For the quark-gluon correlation functions relevant to the leading fermionic pole contribution to the SSAs, we consider the pole structure at $x+y=0$ while $y<0$ since $x>0$ and find that 
\ben
q^2+i\epsilon=2yp^+ q^- -q_\perp^2+i\epsilon=0
\een
provides a pole at
\ben
q^-= - \frac{q_\perp^2}{2|y|p^+}+i\epsilon
\een
in the upper half plane of the $q^-$, while 
\ben
(k+q)^2-m^2+i\epsilon
=2(x+y)p^+(k+q)^{-}-(k_\perp+q_\perp)^2-m^2+i\epsilon
=-(k_\perp+q_\perp)^2-m^2+i\epsilon
\een
does not contribute to any pole in the $q^-$-integration.  That is, when $x+y=0$ and $x>0$, the integrand of $q^-$-integration in Eq.~(\ref{tqf1}) has two poles from $(p-k-q)^2-M_s^2+i\epsilon=0$ and $q^2+i\epsilon=0$ and both of them are in the upper half plane of $q^-$.  Since the integration $dq^-$ in Eq.~(\ref{tqf1}) is sufficiently converging when $|q^-|\to\infty$, the $q^-$ integration vanishes by closing the $q^-$-contour through the lower half plane.  In conclusion, $\tq^{(s)}(0, x)=0$ from this leading order calculation with a scalar diquark, so as $\tq^{(s)}(x, 0)=0$, which can be derived by an explicit calculation or the symmetry property $\tq(x,0)=\tq(0,x)$.

Now we turn to the limit at $y=0$, which is relevant to the leading gluonic pole contribution to the SSAs.  At $y=0$, the pole structure of the $q^-$-integration in Eq.~(\ref{tqf1}) changes.  At $y=0$, 
\ben
(k+q)^2-m^2+i\epsilon=2xp^+q^-
+\frac{x}{1-x}\left[(1-x)M^2-k_\perp^2-M_s^2\right]-(k_\perp^2+q_\perp)^2-m^2+i\epsilon = 0,
\een
leads to a pole at
\ben
q^-=\frac{1}{2xp^+}\left[(k_\perp^2+q_\perp)^2+m^2\right]-\frac{1}{2(1-x)p^+}
\left[(1-x)M^2-k_\perp^2-M_s^2\right]-i\epsilon
\een
in the lower half plane of $q^-$, while 
\ben
q^2+i\epsilon = -q_\perp^2 + i\epsilon
\een
is independent of $q^-$.  Therefore, for the quark-gluon correlation functions relevant to the leading gluonic pole contribution to the SSAs, the integration of $dq^-$ in Eq.~(\ref{tqf1}) has two poles from $(p-k-q)^2-M_s^2+i\epsilon=0$ and $(k+q)^2-m^2+i\epsilon=0$ with one in upper and one in lower half plane of $q^-$.  By closing the $q^-$-contour in either the upper or the lower half plane, we obtain
\ben
\tq^{(s)}(x, x)=\frac{N_c C_F g \lambda_s^2 g_s}{4\pi}(1-x)(m+xM)
\int \frac{d^2k_\perp}{(2\pi)^2}\frac{d^2q_\perp}{(2\pi)^2}
\frac{[q_\perp^2-(q_\perp\cdot s_\perp)^2]\, F(k^2)\, F((k+q)^2)}
{q_\perp^2\left[k_\perp^2+L_s^2(m^2)\right]\left[(k_\perp+q_\perp)^2+L_s^2(m^2)\right]}\, ,
\label{ktqt}
\een
where $k^2$ is given in Eq.~(\ref{k2}) and
\ben
(k+q)^2 = m^2-\frac{1}{1-x}\left[(k_\perp+q_\perp)^2+L_s^2(m^2)\right]
\label{kq2}
\een
with $L_s^2(m^2)$ given in Eq.~(\ref{Ls2}).

The integration over the transverse momenta in Eq.~(\ref{ktqt}) depends on the choice of the form factor.  If we set $F(k^2)=F((k+q)^2)=1$ for the point-like interaction between the nucleon, the constituent quark, and the spectator diquark, we obtain after integrating over $d^2k_\perp$,
\ben
\left.
\tq^{(s)}(x, x)\right|_{\rm point-like}
&=&
\frac{N_c C_F g\lambda_s^2 g_s}{16\pi^2}(1-x)(m+xM)\int_0^1 d\alpha
\int\frac{d^2q_\perp}{(2\pi)^2}\frac{q_\perp^2-(q_\perp\cdot s_\perp)^2}{q_\perp^2
\left[\alpha(1-\alpha)q_\perp^2+L_s^2(m^2)\right]}
\nnu
&=&
\frac{N_c C_F g\lambda_s^2 g_s}{16\pi^2}(1-x)(m+xM)
\int\frac{d^2q_\perp}{(2\pi)^2}
\frac{1}{q_\perp\sqrt{q_\perp^2+4L_s^2(m^2)}}
\ln \frac{\sqrt{q_\perp^2+4L_s^2(m^2)}+q_\perp}
{\sqrt{q_\perp^2+4L_s^2(m^2)}-q_\perp}\, ,
\label{t3_sp}
\een
which has the logarithmic ultraviolet divergence from the region $|q_\perp| \to\infty$.  Since we are interested in the dynamics at the hadronic scale, we could use the dipolar form factor in Eq.~(\ref{dipolar}) to remove the ultraviolet divergence.  Using Eqs.~(\ref{k2}) and (\ref{kq2}), we have 
\ben
F(k^2)  F((k+q)^2) =
(1-x)^2 (\Lambda_s^2)^2
\frac{k_\perp^2+L_s^2(m^2)}{\left[k_\perp^2+L_s^2(\Lambda_s^2)\right]^2}
\frac{(k_\perp+q_\perp)^2+L_s^2(m^2)}{\left[(k_\perp+q_\perp)^2+L_s^2(\Lambda_s^2)\right]^2}\, ,
\label{ff}
\een
thus from Eq.~(\ref{ktqt}),
\ben
\left.\tq^{(s)}(x,  x)\right|_{\rm dipolar}&=&
\frac{N_c C_F g \lambda_s^2 g_s}{8(2\pi)^2}(1-x)^3(m+xM)
\frac{\left(\Lambda_s^2\right)^2}{L_s^2(\Lambda_s^2)}
\int \frac{d^2k_\perp}{(2\pi)^2} \frac{1}{\left[k_\perp^2+L_s^2(\Lambda_s^2)\right]^2}
\nnu
&=&
\frac{N_c C_F g\lambda_s^2 g_s}{16(2\pi)^3}
(1-x)^3(m+xM)\left(\frac{\Lambda_s^2}{L_s^2(\Lambda_s^2)}\right)^2,
\label{tqs}
\een
where $L_s^2(\Lambda_s^2)$ is given in Eq.~(\ref{Ls2}) with $m^2$ replaced by the cutoff scale $\Lambda_s^2$.  Note that the form factor in Eq.~(\ref{ff}) suppresses the ultraviolet contribution to the integration in Eq.~(\ref{ktqt}) without altering the pole structure of the original diagram.  Therefore, our general conclusion on $\tq^{(s)}(0,x)=\tq^{(s)}(x,0)=0$ remains whether we use the dipolar form factor or not. 

The calculation for $\td(x+y, x)$ is identical to that of $\tq(x+y, x)$ except the cut vertex is replaced by the one in Eq.~(\ref{dcut}). Since the pole structure of the diagram is exactly the same, we obtain the same result for correlation functions relevant to the fermionic pole contribution,
\ben
\td^{(s)}(0,x)=-\td^{(s)}(x,0)=0.
\een

From the symmetry property in Eq.~(\ref{tqdq}), we have expected the diagonal correlation function $\td^{(s)}(x, x)$ relevant to the gluonic pole contribution to vanish.  As a consistent test of our model calculation, we verify this result explicitly as follows.  Following the same procedure used to evaluate $\tq^{(s)}(x, x)$ above, we use first the $\delta$-functions to integrate over $k^+, k^-, q^+$, then the pole structure to integrate over $q^-$ to get the necessary phase, and we obtain
\ben
\td^{(s)}(x, x)=\frac{N_c C_F g\lambda_s^2 g_s}{4\pi}
(1-x)(m+xM)\int \frac{d^2k_\perp}{(2\pi)^2}\frac{d^2q_\perp}{(2\pi)^2}
\frac{q_\perp\cdot s_\perp [2k_\perp\cdot s_\perp+q_\perp\cdot s_\perp]F(k^2)\, F((k+q)^2)}
{q_\perp^2[k_\perp^2+L_s^2(m^2)][(k_\perp+q_\perp)^2+L_s^2(m^2)]}\, .
\een
We first consider the point-like interaction case setting $F(k^2) F((k+q)^2)=1$.  Using the Feynman parametrization to combine the $k_\perp$ dependent denominator, we obtain
\ben
\left.\td^{(s)}(x, x)\right|_{\rm point-like}
&=&\frac{N_c C_F g\lambda_s^2 g_s}{4\pi}
(1-x)(m+xM)\int \frac{d^2q_\perp}{(2\pi)^2} \frac{d^2\ell_\perp}{(2\pi)^2}\int_0^1 d\alpha
\frac{(1-2\alpha)(q_\perp\cdot s_\perp)^2}{q_\perp^2\left[\ell_\perp^2+\alpha(1-\alpha)q_\perp^2+L_s^2(m^2)\right]^2}
\nonumber \\
&=&
0\, .
\een
Here, the second line is due to the fact that the numerator of the $\alpha$ integral is antisymmetric under $\alpha\leftrightarrow 1-\alpha$ while the denominator and the integration limits are symmetric.  From Eq.~(\ref{ff}), it is clear that the inclusion of the dipolar form factor does not change the main feature of the $\alpha$-dependence of the combined denominator, 
\ben
\left.\td^{(s)}(x, x)\right|_{\rm dipolar}
=\frac{N_c C_F g\lambda_s^2 g_s}{4\pi}
(1-x)^3(m+xM)(\Lambda_s^2)^2
\int \frac{d^2q_\perp}{(2\pi)^2} \frac{d^2\ell_\perp}{(2\pi)^2}\int_0^1 d\alpha\,
\frac{3!\, \alpha(1-\alpha)(1-2\alpha)(q_\perp\cdot s_\perp)^2}{q_\perp^2\left[\ell_\perp^2+\alpha(1-\alpha)q_\perp^2+L_s^2(m^2)\right]^4}
\een
which also vanishes from the symmetry of the $d\alpha$ integration.  We thus verify that $\td^{(s)}(x, x)=0$.
\bef
\psfig{file=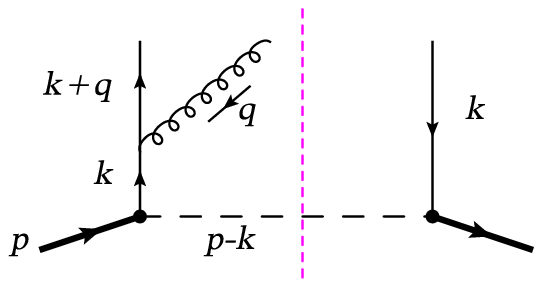, width=2.1in}
\caption{Feynman diagram at the first non-trivial order that could potentially contribute to the twist-3 quark-gluon correlation functions.}
\label{extra}
\eef

To conclude this subsection on the calculation with a scalar diquark, we make a comment on the contribution to the quark-gluon correlation function from the diagram in Fig.~\ref{extra}.  In order to get the SSAs, as discussed earlier, we need a spin flip between the two partonic states on the opposite side of the cut in the diagram in Fig.~\ref{extra}.  Since the quark-gluon composite state on the left was initiated from a single quark state, the spin flip contribution can only come from the mass term of the quark.  Therefore, the contribution of the diagram in Fig.~\ref{extra} to the quark-gluon correlation functions relevant to the SSAs is expected to be proportional to the quark mass and is therefore small.  Our explicit calculation shows that the diagonal correlation function $\tq(x, x)$ relevant to the leading gluonic pole contribution vanishes at this order which is consistent with the fact that when the gluon momentum vanishes, there is no spin flip between two quarks on the opposite side of the cut.  For the off-diagonal correlation functions relevant to the fermionic pole contribution, we find 
\ben
\left. \tq^{(s)}(0, x)\right|_{\rm Fig.~\ref{extra}}=
-\left. \td^{(s)}(0, x)\right|_{\rm Fig.~\ref{extra}}&=&
\frac{N_c C_Fg^2\lambda_s^2}{8\pi}(1-x)m (m+xM)^2
\nnu
&\ &
\times
\int \frac{d^2k_\perp}{(2\pi)^2}\frac{d^2q_\perp}{(2\pi)^2}
\frac{F(k^2)^2}{\left[q_\perp^2+m^2\right]\left[k_\perp^2+L_s^2(m^2)\right]^2}\, ,
\een
which is clearly proportional to the mass of quark.

%%%%%%%%%%%%%%%%%%%%%%
\subsection{Calculation with an axial-vector diquark}

In order to test the sensitivity of our results derived in the last subsection on the choice of the scalar diquark, we present in this subsection quark-gluon correlation functions calculated with an axial-vector diquark.

By using the same Feynman diagram in Fig.~\ref{tffey} with the Feynman rule for an axial-vector spectator, and the same cut vertices, we derive the quark-gluon correlation functions relevant to both leading gluonic and fermionic pole contribution to the SSAs.  Since the pole structure of the Feynman diagram in Fig.~\ref{tffey} is insensitive to whether the spectator is a scalar or an axial-vector, we find, like the case of a scalar diquark, that all off-diagonal quark-gluon correlation functions relevant to the leading fermionic pole contribution vanish,
\ben
\tq^{(v)}(0, x)=\td^{(v)}(0,x)=0.
\een
For the diagonal quark-gluon correlation functions relevant to the leading gluonic pole contribution, we obtain
\ben
\tq^{(v)}(x, x)&=&
\frac{N_c C_F g\lambda_v^2 g_v}{4\pi}\, x(m+xM)
\int \frac{d^2k_\perp}{(2\pi)^2}\frac{d^2q_\perp}{(2\pi)^2}
\frac{[q_\perp^2-(q_\perp\cdot s_\perp)^2]\, F(k^2) F((k+q)^2)}
{q_\perp^2\left[k_\perp^2+L_s^2(m^2)\right]
\left[(k_\perp+q_\perp)^2+L_s^2(m^2)\right]}\, ,
\label{ktqtv}
\een
which is the same as that in Eq.~(\ref{ktqt}) except the overall $(1-x)$ factor is replaced by $x$ due to the difference in diquark spin. Therefore, the rest of derivation and discussion in the last subsection following Eq.~(\ref{ktqt}) should be the same for the case of an axial-vector diquark. We find
\ben
&&\left. \tq^{(v)}(x, x)\right|_{\rm point-like}
=
\frac{N_c C_F g\lambda_v^2 g_v}{16\pi^2}x(m+xM)
\int\frac{d^2q_\perp}{(2\pi)^2}
\frac{1}{q_\perp\sqrt{q_\perp^2+4L_s^2(m^2)}}
\ln \frac{\sqrt{q_\perp^2+4L_s^2(m^2)}+q_\perp}
{\sqrt{q_\perp^2+4L_s^2(m^2)}-q_\perp}\, ,
\\
&&\left. \tq^{(v)}(x, x)\right|_{\rm dipolar}
=\frac{N_c C_F g\lambda_v^2 g_v}{16(2\pi)^3}\, x(1-x)^2(m+xM)\left(\frac{\Lambda_s^2}{L_s^2(\Lambda_s^2)}\right)^2\, ,
\label{tqv}
\een
which are the same as those in Eqs.~(\ref{t3_sp}) and (\ref{tqs}) except that one factor of $(1-x)$ is replaced by $x$.  We also explicitly verify that $\td(x, x)=0$ when it is calculated with an axial-vector diquark. 

To conclude this section, we summarize our key results as follows.  We find, in terms of an explicit calculation in the quark-diquark model of the nucleon, that at the leading non-trivial order all quark-gluon correlation functions relevant to the leading fermionic pole contribution to the SSAs vanish,
\ben
\tq(0, x)=\tq(x,0)=0, \qquad
\td(0, x)=-\td(x,0)=0\, .
\een
We also verify that $\td(x,x)=0$, and find that only the diagonal quark-gluon correlation function, $\tq(x,x)$, is finite.

%%%%%%%%%%%%%%%%%%%%%%
\section{Connection to TMD parton distribution functions}
\label{conn}

As we stressed in the introduction of this paper, the collinear factorization approach and the
TMD factorization approach to the SSAs are closely connected and complementary to each other.
It was shown in terms of their operator definitions that the twist-3 quark-gluon correlation
function $\tq(x, x)$ is related to the moment of quark Sivers function $f_{1T}^\perp(x, k_\perp^2)$ \cite{boermulders},
\ben
\tq(x, x)=\frac{1}{M}\int d^2k_\perp k_\perp^2 f_{1T}^\perp(x, k_\perp^2),
\label{t3sivers}
\een
where $f_{1T}^\perp(x, k_\perp^2)$ is the quark Sivers function defined via the Drell-Yan process, which is related to the quark Sivers function defined in the semi-inclusive deep inelastic scattering by a minus sign \cite{Collins:2002kn,Kang:2009bp}.  In this section, we explicitly verify this relation in Eq.~(\ref{t3sivers}) by comparing the quark-gluon correlation function $\tq(x,x)$ calculated in this paper with the Sivers function calculated in the same quark-diquark model of the nucleon.

\bef
\psfig{file=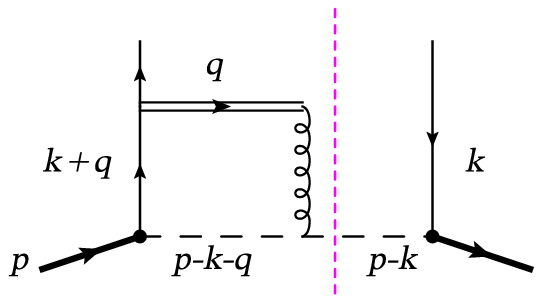, width=2.1in}
\caption{Lowest order Feynman diagram for the quark Sivers function in the quark-diquark model of the nucleon \cite{Bacchetta:2008af}.}
\label{siversfey}
\eef
The quark Sivers function in the quark-diquark model of the nucleon has been calculated in Ref.~\cite{Bacchetta:2008af}. The Feynman diagram to the lowest non-trivial order for the quark Sivers function is shown in Fig.~\ref{siversfey}. In terms of an explicit calculation with a scalar diquark we obtain 
\ben
\left. f_{1T}^{\perp(s)}(x, k_\perp^2)\right|_{\rm point-like}
&=&
\frac{N_c C_Fg\lambda_s^2 g_s}{4(2\pi)^4}
\frac{(1-x)M(m+xM)}{k_\perp^2\left[k_\perp^2+L_s^2(m^2)\right]}
\ln\frac{k_\perp^2+L_s^2(m^2)}{L_s^2(m^2)},
\label{sivers_sp}
\een
for a point-like interaction between the nucleon, the quark, and the spectator diquark, and
\ben
\left.f_{1T}^{\perp(s)}(x, k_\perp^2)\right|_{\rm dipolar}
&=&
\frac{N_c C_Fg\lambda_s^2 g_s}{4(2\pi)^4}\,
(1-x)^3M(m+xM)
\frac{[\Lambda_s^2]^2}{L_s^2(\Lambda_s^2)\left[k_\perp^2+L_s^2(\Lambda_s^2)\right]^3} 
\label{sivers_sf}
\een
for using the dipolar form factor for the interaction between the nucleon, the quark, and the spectator diquark.  Similarly, we find, in terms of the calculation with an axial-vector diquark,
\ben
\left. f_{1T}^{\perp(v)}(x, k_\perp^2)\right|_{\rm point-like}
&=&\frac{N_c C_Fg\lambda_v^2 g_v}{4(2\pi)^4}
\frac{xM(m+xM)}{k_\perp^2\left[k_\perp^2+L_s^2(m^2)\right]}
\ln\frac{k_\perp^2+L_s^2(m^2)}{L_s^2(m^2)},
\label{sivers_vp}
\een
and
\ben
\left.f_{1T}^{\perp(v)}(x, k_\perp^2)\right|_{\rm dipolar}
&=&\frac{N_c C_Fg\lambda_v^2 g_v}{4(2\pi)^4}\,
x(1-x)^2 M(m+xM)
\frac{[\Lambda_s^2]^2}{L_s^2(\Lambda_s^2)\left[k_\perp^2+L_s^2(\Lambda_s^2)\right]^3},
\label{sivers_vf}
\een
respectively.  The calculated results here are the same as those obtained in Ref.~\cite{Bacchetta:2008af} except the overall constant factor $[\Lambda_s^2]^2$ for those with the dipolar form factor. The difference is, as explained earlier, due to a slightly different choice of the form factor so that the calculated twist-3 correlation functions as well as the quark Sivers function appear to have the right dimension. 

In order to verify the relation in Eq.~(\ref{t3sivers}), we need to take the moment of the quark Sivers functions calculated above in the same quark-diquark model of the nucleon.  However, the moment of the quark Sivers functions in Eqs.~(\ref{sivers_sp}) and (\ref{sivers_vp}) calculated by using the point-like interaction is logarithmically divergent, for example, 
\ben
\frac{1}{M}\int d^2k_\perp k_\perp^2 \left. f_{1T}^{\perp(s)}(x, k_\perp^2)\right|_{\rm point-like}
&=&
\frac{N_c C_Fg\lambda_s^2 g_s}{16\pi^2}\,
(1-x)(m+xM) 
\nonumber\\
&\ &\times
\int \frac{d^2k_\perp}{(2\pi)^2}
\frac{1}{\left[k_\perp^2+L_s^2(m^2)\right]}
\ln\frac{k_\perp^2+L_s^2(m^2)}{L_s^2(m^2)}\, .
\label{sivers_ms}
\een
The moment of the quark Sivers function in Eq.~(\ref{sivers_ms}) is clearly not necessary to be the same as the twist-3 quark-gluon correlation function in Eq.~(\ref{t3_sp}), even if one imposes the same ultraviolet cutoff on the transverse momentum integration in both Eqs.~(\ref{t3_sp}) and (\ref{sivers_ms}).  This is because the ultraviolet divergence of the calculated quark-gluon correlation function in Eq.~(\ref{t3_sp}) was not regularized in the same way as that in the Sivers function calculation in Eq.~(\ref{sivers_sp}).  This example indicates that if one wants to compare the both sides of Eq.~(\ref{t3sivers}) {\it perturbatively} by projecting the equation onto a parton state, one has to specify the regularization and renormalization condition for the ultraviolet divergence in both sides.  In general, the relation in Eq.~(\ref{t3sivers}) is not necessarily valid for all orders in perturbative calculations if one does not regularize and renormalize the ultraviolet divergence in the same way for the both sides. 

If we regularize and renormalize the ultraviolet divergence in both sides of Eq.~(\ref{t3sivers}) in the same way, we should expect the relation to hold.  To explicitly demonstrate this, we compare the quark-gluon correlation functions and the quark Sivers function calculated with the same dipolar form factor, 
\ben
&\mbox{scalar diquark:}&\qquad
\frac{1}{M}\int d^2k_\perp k_\perp^2 f_{1T}^{\perp(s)}(x, k_\perp^2)
=\frac{N_c C_F g\lambda_s^2 g_s}{16(2\pi)^3}
(1-x)^3(m+xM)\left[\frac{\Lambda_s^2}{L_s^2(\Lambda_s^2)}\right]^2,
\\
&\mbox{axial-vector diquark:}&\qquad
\frac{1}{M}\int d^2k_\perp k_\perp^2 f_{1T}^{\perp(v)}(x, k_\perp^2)
=\frac{N_c C_F g\lambda_v^2 g_v}{16(2\pi)^3}\,
x(1-x)^2(m+xM)\left[\frac{\Lambda_s^2}{L_s^2(\Lambda_s^2)}\right]^2.
\een
The right-hand-side of above equations are clearly equal to the quark-gluon correlation functions in Eqs.~(\ref{tqs}) and (\ref{tqv}), respectively.  

%%%%%%%%%%%%%%%%%%%%%%
\section{Summary and conclusions}
\label{sum}

In this paper, we calculate various twist-3 quark-gluon correlation functions of a transversely polarized nucleon in the quark-diquark model of the nucleon.  Our calculations are done with the diquark being a scalar particle as well as being an axial-vector particle.  We have found from our calculation at the first non-trivial order that all quark-gluon correlation functions relevant to the leading fermionic pole contribution, $\tq(0, x)$, $\td(0, x)$, $\tq(x,0)$, and $\td(x,0)$, vanish.  Only one of the diagonal quark-gluon correlation functions relevant to the leading gluonic pole contribution, $\tq(x,x)$, is finite.  The other diagonal quark-gluon correlation function, $\td(x,x)$, also vanishes from both the symmetry argument and explicit calculation. Our conclusions are independent of the diquark being a scalar or an axial-vector.  

Although our results are derived from a specific model calculation, the features of the calculated results should allow us to conclude with  confidence that the diagonal quark-gluon correlation function $\tq(x,x)$ is much larger than all other quark-gluon correlation functions that are relevant to the leading soft pole contribution to the SSAs. This conclusion is significant and important for phenomenological study of the SSAs. It enables us to study the physics of SSAs without including too many unknown correlation functions at the early stage of probing this new domain of QCD dynamics.  However, it is the limitation of the quark-diquark model of the nucleon that we are not able to calculate the tri-gluon correlation functions in this model, which are closely connected to the quark-gluon correlation functions via perturbative evolution \cite{kqevo,Braun:2009mi}.

As we explained in last section, it requires a caution in using the relation between the twist-3 quark-gluon correlation function $\tq(x,x)$ and the moment of the quark Sivers function in Eq.~(\ref{t3sivers}). Since both sides of the equation, the twist-3 quark-gluon correlation function $\tq(x,x)$ on the left and the moment of quark Sivers function on the right, are perturbatively divergent, the relation is valid only if the same regularization and renormalization scheme is adopted to the calculation of both sides. As an example, we demonstrate in the last section that the relation could be violated perturbatively if different regularization and renormalization schemes were used; and the relation is valid if the same scheme were used in both sides.

%%%%%%%%%%%%%%%%%%%%%%
\section*{Acknowledgments}
We thank M.~Burkardt and G.~Sterman for helpful discussions.  This work was supported in part by the U. S. Department of Energy under Grant No.~DE-FG02-87ER40371. Z.K. and J.Q. are grateful to RIKEN/BNL Research Center, Brookhaven National Laboratory, and the U.S. Department of Energy (Contract No.~DE-AC02-98CH10886) for providing the support and facilities essential for the completion of this work. 

%%%%%%%%%%%%%%

\end{document}